\documentclass[a4paper,11pt]{article}
\pdfoutput=1

\usepackage{amsmath}
\usepackage{amssymb}
\usepackage[small]{caption}
\usepackage[margin=1in]{geometry}
\usepackage[multiple]{footmisc}
\usepackage{graphicx}
\usepackage{mathtools}
\usepackage{physics}
\usepackage{slashed}
\usepackage{xcolor}

\definecolor{mediumblue}{rgb}{0,0,0.8}
\usepackage{hyperref}
\hypersetup{
  linktocpage=true,
  colorlinks=true,
  citecolor=mediumblue,
  filecolor=mediumblue,
  linkcolor=mediumblue,
  urlcolor=mediumblue,
}
\newcommand{\mailref}[1]{\href{mailto:#1}{#1}}

\graphicspath{{./}{./figures/}}
\allowdisplaybreaks%

\newcommand{\diag}{\mathop{\mathrm{diag}}\nolimits}
\newcommand{\CC}{C\nolinebreak\hspace{-.05em}\raisebox{.4ex}{\tiny\bf +}\nolinebreak\hspace{-.10em}\raisebox{.4ex}{\tiny\bf +}}

\def\thefootnote{\fnsymbol{footnote}}

\begin{document}

\begin{titlepage}
  \begin{flushright}
    CTPU-PTC-20-14
  \end{flushright}

  \bigskip

  \begin{center}
    \bf \LARGE A singular way to search for heavy resonances\\[2mm]
    in missing energy events
  \end{center}

  \medskip

  \begin{center}
    \bf \large Chan~Beom~Park\footnote{\texttt{\mailref{cbpark@ibs.re.kr}}}
  \end{center}

  \begin{center}
    \em Center for Theoretical Physics of the Universe,\\
    Institute for Basic Science (IBS), Daejeon 34126, Korea\\[0.2cm]
  \end{center}

  \medskip

  \begin{abstract}
    The phase space of visible particles in missing energy events may
    have singularity structures. The singularity variables are devised
    to capture the singularities effectively for given event
    topology. They can greatly improve the discovery potential of new
    physics signals as well as to extract the mass spectrum information
    at hadron colliders.
    Focusing on the antler decay topology of resonance, we derive a
    novel singularity variable whose distribution has endpoints
    directly correlated with the resonance mass.
    As a practical application, we examine the applicability of the
    singularity variable to the searches for heavy neutral Higgs
    bosons in the two-Higgs doublet model.
  \end{abstract}
\end{titlepage}

\renewcommand{\thefootnote}{\arabic{footnote}}
\setcounter{footnote}{0}

\section{Introduction}

\noindent
Signals with missing energy at hadron colliders commonly arise in many
new physics models solving the dark matter problem of the Universe.
Even in the Standard Model (SM), the neutrinos are undetectable, so
recorded as missing energy.
Though being ubiquitous, the missing energy has been a
challenging object that prevents the full reconstruction of particles
involved in the decay process.
As will be discussed in Sec.~\ref{sec:singularity}, it is because of
the fundamental absence of inverse projection from the phase space of
visible particle momenta to the full phase space.
Still, there are a plethora of useful methods and algorithms proposed
for determining the mass spectrum of the underlying dynamics from
missing energy events~\cite{Barr:2010zj, Barr:2011xt}.

A mathematical insight on the missing energy kinematics led to the
invention of the algebraic kinematic method~\cite{Kim:2009si}.
It was realized that the phase space of visible particles might
possess identifiable singularities, from which one could extract the
mass spectrum information for given event topology.
Then, an optimized one-dimension variable called the singularity
coordinate has been proposed to capture the singular behavior in an
effective way.
The distribution of the singularity coordinate becomes singular, {\em
  i.e.}, having a sharp peak or distinct edge, when the input masses
equal to the true values.
Though powerful and insightful, it has not been widely considered as
applicable for practical new physics searches, partly due to the lack
of concrete prescriptions with more examples.
Another obstructing factor is that the singularity coordinate is
an implicit function of the mass spectrum involved in the decay
process.
In the absence of good ansatz, it is necessary to perform
multi-dimensional fitting, which is impractical at the stage of
discovery.
Furthermore, as the value of the singularity coordinate does not
relate directly to physical parameters, it is not trivial to interpret
the singularity coordinate in terms of the physical quantities.
The situation has recently improved due to the studies in
Ref.~\cite{Matchev:2019bon}, where the singularity method is
reexamined and expanded for various event topologies.

Motivated by the results in Ref.~\cite{Matchev:2019bon}, we examine
the singularity method for the antler decay topology~\cite{Han:2009ss,
  Han:2012nm, Han:2012nr}, and propose a derived singularity variable.
Many new physics resonances decaying into the final state with missing
energy can be represented by the antler decay topology.
For example, the heavy neutral Higgs boson in the supersymmetric model
may decay into the final state with invisible neutralinos:
$H \to \widetilde\chi_2^0 +
\widetilde\chi_2^0 \to Z\widetilde\chi_1^0 + Z
\widetilde\chi_1^0$~\cite{Djouadi:2005gj}.
In the two-Higgs doublet model (2HDM), one of the most important decay
modes of the heavy Higgs bosons is $H/A \to t + \bar t \to b \ell^+ \nu
+ \bar b \ell^- \bar\nu$~\cite{Branco:2011iw}.
The same final state can arise from the decay of extra $U(1)$ gauge
boson, $Z^\prime \to t + \bar t$~\cite{Langacker:2008yv}.
The doubly-charged scalar boson $H^{\pm\pm}$ in the Higgs triplet
model may decay into same-sign $W$ boson pairs: $H^{++} \to W^+ + W^+
\to \ell^+ \nu + \ell^+ \nu$~\cite{Kanemura:2014goa}, which
corresponds to the antler decay topology.

The novel singularity variable has a direct correlation with the mass
scale of the resonance, so it can greatly help distinguish the new
physics signal from backgrounds.
And, along with the inclusion of supplementary approximation for the
unknown longitudinal momentum of the resonance, the endpoint shape of
the singularity variable distributions becomes more pronounced, so we
expect that it can be useful for the mass measurement as well.
A brief overview of the singularity method and the derivation of the
new singularity variable are presented in the next section.

In Sec.~\ref{sec:heavy_higgs}, we evaluate the performance of the
singularity variable in the case of a heavy Higgs boson decaying to a
top pair.
We compare the signal distributions to the dominant SM backgrounds and
take the detector effects into account, to check the viability of the
singularity variable in more realistic collider searches.
Since the unknown longitudinal momentum of the Higgs boson produced at
hadron colliders is not negligible, we improve the singularity
variable by employing an approximation scheme.

\section{\label{sec:singularity}
  Kinematic singularity of the antler decay topology}

\noindent
The phase space is the hypersurface of the final-state particle
momenta, subject to the kinematic constraints like energy-momentum
conservation and on-shell mass relations.
In other words, the particle momenta reside in the solution space of
coupled polynomial equations of total degree two.
The set of all solutions of a system of polynomial equations is called
an {\em affine variety} in mathematics:
\begin{equation}
  \Pi(g_1, \, \dots, \, g_m)
  = \left\{ (P_1, \, \dots, \, P_n) \in \mathbb{E}^n \,\vert\,
    g_i (P_1, \, \dots, \, P_n) = 0 \,\,\text{for all}\,\, 1 \leq i
    \leq m \right\} ,
  \label{eq:affine_variety}
\end{equation}
where $P_i$ are the four-momenta of final state particles, and $g_i$
are polynomials for kinematic constraints such as on-shell mass
relations. $\mathbb{E}$ is the four-dimensional pseudo-Euclidean space,
$\mathbb{R}^{1, \, 3}$.

We often confront the situation where we cannot fully reconstruct the
phase space.
It is because the final states of collider events may contain invisible
particles such as neutrinos or dark matter candidates, which escape
human-made detectors, as well as visible particles.
The missing energy events are one of the typical types of signals
predicted by many new physics models beyond the SM that we eager to
discover at the LHC and future colliders.
We decompose the phase space into $\{ (p_i, \, k_j)\}$, where $p_i$ are
the four-momenta of visible particles, and $k_j$ are those of invisible
particles.
In terms of the phase space, what detectors at collider experiments
are doing is the projection of the full phase space $\{ (p_i, \, k_j)\}$
onto the space of visible momenta $\{ (p_i) \}$, up to finite detector
resolution and acceptance, for the missing energy events.
The projection makes the reconstruction of the center-of-mass frame
hard, or even impossible, on an event-by-event basis because it is not
invertible.
The only invertible projection is the identity mapping, which is
unachievable by construction.

One important property of the affine variety is the presence of
singular points or {\em singularities}.
In Ref.~\cite{Kim:2009si}, it was noted that the projected visible
phase space could possess singularities, where the tangent plane fails
to exist, even though the full phase space including the invisible
momenta was regular at the points.
To formulate it, we introduce the Jacobian matrix $J$, the $m \times
n$ matrix of partial derivatives:
\begin{equation}
  J_{ij} = \pdv{g_i}{P_j},
\end{equation}
where $g_i$ are the defining polynomial of the phase space in
(\ref{eq:affine_variety}).
If the point $Q$ is regular, the Jacobian matrix provides the linear
approximation of the phase space near the point.
It is the tangent space at $Q$,
\begin{equation}
  \sum_{j=1}^n \frac{\partial g_i}{\partial P_j} (P_j - Q_j) \quad
  (i = 1, \, \dots,\, m).
\end{equation}
The rank of $J$ equals the dimension of the phase space.
At the singularity, the Jacobian matrix has a reduced rank, lower than
the rank at the regular point.
In the case where $m = n$, the determinant of the Jacobian matrix
vanishes at the singularity.
In Ref.~\cite{Kim:2009si}, it is considered the restricted matrix
composed of the derivatives for the invisible momenta, rather than the
full Jacobian matrix. Then, a singularity coordinate has been proposed
to exploit the singularity structure of the visible phase
space.\footnote{The singularity appears as edge or cusp in the
  visible phase space, depending on the amount of the reduced rank of
  the Jacobian matrix.}
It is an implicit function of the mass spectrum involved in the
hypothesized decay process for given events.
Once the hypothesis was correct, the singularity coordinates maximize
the singular features at the true mass values, thus enabling us to
determine the mass spectrum.

Though being a simple and elegant idea based on mathematical
constructions, one has failed to find practical examples except for the
applications to the $W \to \ell \nu$~\cite{Rujula:2011qn} and $h \to
W W \to 2 \ell + \slashed{E}_T$ processes~\cite{DeRujula:2012ns} at
hadron colliders.
It is due to the lack of more concrete examples and programmed
implementations for practitioners.
Recently, a set of worked-out examples for various event topologies
have been thoroughly investigated and visualized in
Ref.~\cite{Matchev:2019bon}, which deepens the understanding of the
singularity variables and makes them more approachable.

\begin{figure}[tb!]
  \begin{center}
    \includegraphics[width=0.48\textwidth]{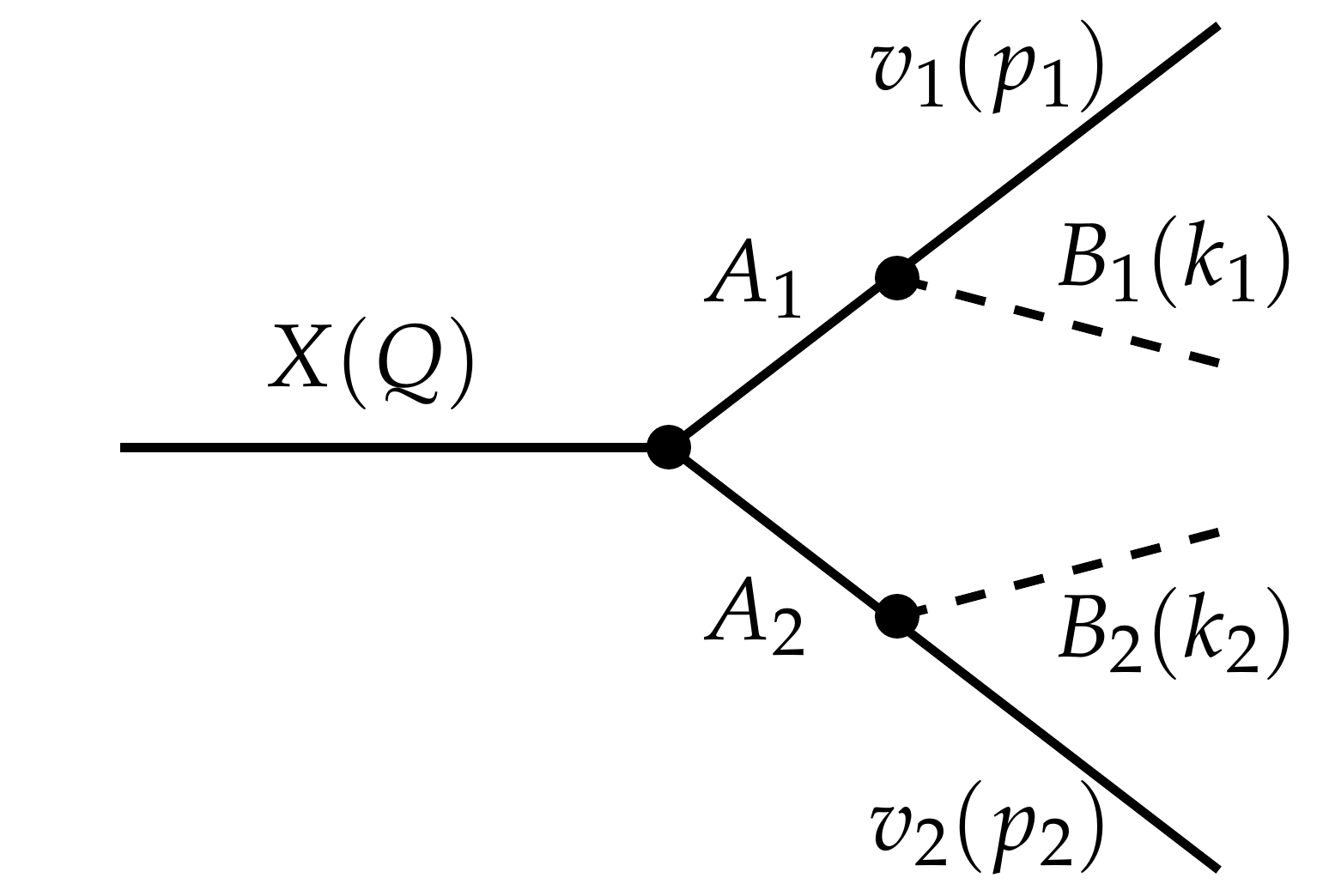}
  \end{center}
  \caption{\label{fig:antler_diagram}
  The antler decay topology. A heavy resonance ($X$) decays to a pair of
  lighter intermediate particles ($A_{1,\, 2}$), and they subsequently
  decays into a pair of visible ($v_{1, \, 2}$) and invisible
  particles ($B_{1,\, 2}$).}
\end{figure}

Inspired by the studies in~\cite{Matchev:2019bon}, we here concentrate
our attention on the antler decay topology and propose a novel
kinematic variable derived from the singularity variable.
In the antler decay diagram, a singly produced heavy resonance decays
into a pair of visible particles and a pair of invisible particles via
unstable intermediate states~\cite{Han:2009ss, Han:2012nm,
  Han:2012nr}, as shown in Fig.~\ref{fig:antler_diagram}.
It represents a typical decay pattern of heavy resonances in many new
physics models such as the decays of the heavy Higgs bosons in the
2HDM, as well as the
$h \to W W \to 2 \ell + \slashed{E}_T$ process in the SM\@.
A technical subtlety of the singularity coordinate is that it is an
implicit variable whose relation with the mass spectrum cannot
directly be inferred.
It differs from invariant and transverse mass variables, whose peak or
edge directly provides the mass information.
On the other hand, the peak position of the singularity coordinate is
not related to the mass value.
What the singularity coordinate promises is that the distribution
exhibits singular behavior when the hypothesis on the mass spectrum is
correct.
One can attempt the maximum likelihood estimation for the mass
spectrum using template distributions, but it is not suitable at the
stage of discovery.
It is clear that kinematic variables would be more useful if they
enable us to directly deduce the mass scale of heavy resonances by
identifying the positions of the peak or edge of their distributions.
To derive such a kinematic variable, we review the singularity
variable for the antler decay topology at first.

As shown in Fig.~\ref{fig:antler_diagram}, we consider a heavy
resonance $X$, decaying into two visible $v_{1, \, 2}$ and two
invisible particles $B_{1, \, 2}$ through unstable intermediate states
$A_{1, \, 2}$,
\begin{equation}
  X (Q) \longrightarrow A_1 + A_2
  \longrightarrow v_1 (p_1) \, B_1 (k_1) + v_2 (p_2) \, B_2 (k_2) .
  \label{eq:antler}
\end{equation}
Notice that the diagram also covers the case of a pair of cascade
decay chains. One can define $v_i$ to be the collection of visible
particles at the chain $i$. In this case, the invariant mass of the
visible particle system $m_i$ is not constant, but an event variable.
We take into account the case where all the particles
are on mass-shell and the decay chains to be symmetric: $M_{A_1} =
M_{A_2} = M_A$ and $M_{B_1} = M_{B_2} = M_{B}$.
$B_i$ corresponds to the neutrino or a dark matter candidate, yielding
the missing transverse energy,
\begin{equation}
  \vb*{k}_{1T} + \vb*{k}_{2T} = \slashed{\vb*{P}}_T .
\end{equation}
In the cases where the intermediate states are off mass-shell, or
their decay widths are large, the on-shell mass relations are not
valid.
It results in a decreased number of kinematic constraints.
While the singularity method is still valid, we may not
use the condition of vanishing Jacobian matrix, but have to find the
reduced rank condition of the Jacobian.

Using the condition of energy-momentum conservation
$Q = p_1 + p_2 + k_1 + k_2$,
the kinematic constraints for the antler decay topology
(\ref{eq:antler}) are given as:
% \begin{align}
%   \begin{split}
%     k_1^2
%     &= M_B^2 , \\
%     2 p_1 \cdot k_1
%     &= M_A^2 - M_B^2 - m_1^2, \\
%     2 p_2 \cdot k_1
%     &= - M_A^2 + M_B^2 + 2(Q - p_1) \cdot p_2 - m_2^2
%     , \\
%     2 Q \cdot k_1
%     &= Q^2 - 2 Q \cdot p_1 .
%   \end{split}
% \end{align}
\begin{align}
  \begin{split}
    g_1 &= k_1^2 - M_B^2 , \\
    g_2 &= 2 p_1 \cdot k_1 - M_A^2 + M_B^2 + m_1^2, \\
    g_3 &= 2 p_2 \cdot k_1 - 2 Q \cdot p_2 + 2 p_1 \cdot p_2
     + M_A^2 - M_B^2 + m_2^2 , \\
    g_4 &= 2 Q \cdot k_1 + 2 Q \cdot p_1 - M_X^2 .
  \end{split} \label{eq:kinematic_constraints}
\end{align}
And, the elements of the Jacobian matrix are obtained by taking
partial derivatives on the constraint equations,
\begin{equation}
  J_{ij} = \frac{\partial g_i}{\partial k_{1j}}
\end{equation}
for $k_{1j} = (k_{10},\, \vb*{k}_{1T},\, k_{1L})$.
Since it is a square matrix, the singularity condition implies that
its determinant is zero.
To eliminate the invisible momentum components, we adopt a simple
trick used in Ref.~\cite{Matchev:2019bon},
where $\det (J \eta J^\mathsf{T})$ with
$\eta = \diag (1, \, -1, \, -1, \, -1)$ is taken
instead of $\det J$.
Then, we find that the singularity condition is equivalent to the
vanishing of\footnote{Here we have performed some row and column
  operations to simplify the expression.}
\begin{equation}
  \Delta_\text{AT}
  \equiv
  \text{
  {\footnotesize
  $\begin{vmatrix}
    2 M_A^2
    & M_A^2 - M_B^2 + m_1^2
    & 2 Q \cdot p_2 - M_A^2 + M_B^2 - m_2^2
    & M_X^2 \\
    M_A^2 - M_B^2 + m_1^2
    & 2 m_1^2 & 2 p_1 \cdot p_2 & 2 Q \cdot p_1 \\
    2 Q \cdot p_2 - M_A^2 + M_B^2 - m_2^2
    & 2 p_1 \cdot p_2 & 2 m_2^2 & 2 Q \cdot p_2 \\
    M_X^2 & 2 Q \cdot p_1 & 2 Q \cdot p_2 & 2M_X^2
  \end{vmatrix}$}} .
\end{equation}
Here the subscript ``AT'' stands for the antler decay topology.
Note that the transverse momentum of the heavy resonance can be
determined by measured quantities because
\begin{equation}
 \vb*{Q}_T = \vb*{p}_{1T} + \vb*{p}_{2T} + \slashed{\vb*{P}}_T .
\end{equation}
Thus, $\Delta_\text{AT}$ is a function of $Q_0 = (M_X^2 +
  \norm{\vb*{Q}_T}^2 + Q_L^2)^{1/2}$ and $Q_L$ if $M_A$ and $M_B$
were known {\em a priori} or determined by ansatz.
Given $Q_L$, $\Delta_\text{AT}$ is a quartic polynomial equation of
$Q_0$ with nonzero coefficients, in general.
If the resonance was produced at rest, {\em i.e.}, $Q_0 = M_X$ and
$Q_L = 0$, $\Delta_\text{AT}$ is a quartic equation of $M_X$ in the
form like $M_X^2 (a M_X^2 + b M_X + c)$.
The trivial solution $M_X = 0$ is the side effect of the determinant
trick, which is unphysical. Therefore, $\Delta_\text{AT}$ reduces to a
quadratic equation of $M_X$ in essence.

\begin{figure}[tb!]
  \begin{center}
    \includegraphics[width=0.48\textwidth]{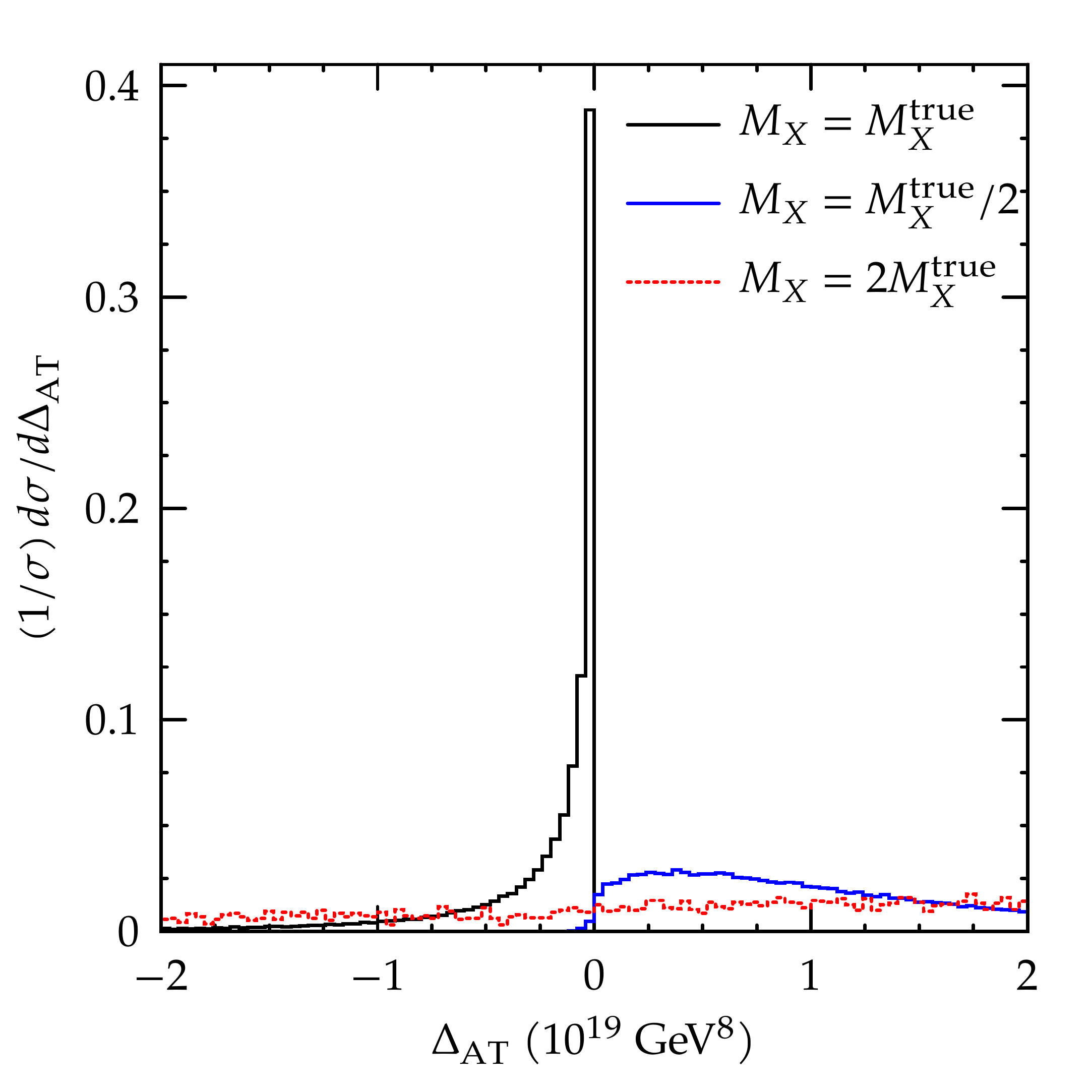}
    \includegraphics[width=0.48\textwidth]{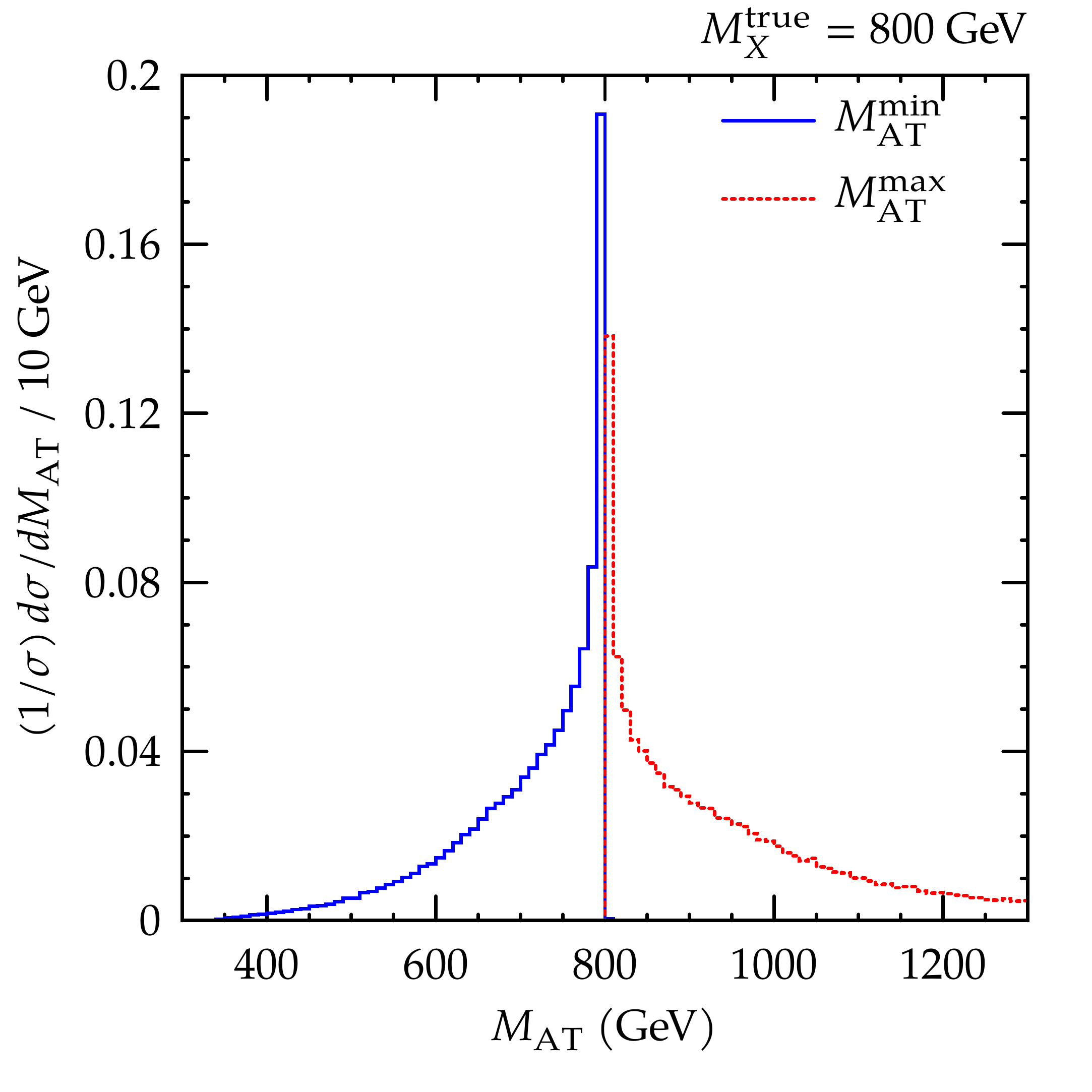}
  \end{center}
  \caption{\label{fig:at_ps}
    The $\Delta_\text{AT}$ (left) and $M_\text{AT}$ (right)
    distributions for pure phase space.
    % We have used the true spatial momentum of the resonance.
    The resonance $X$ is produced at rest, $\vb*{Q}_T = \mathbf{0}$
    and $Q_L = 0$.
    For $\Delta_\text{AT}$, the resonance
    mass $M_X$ is the input parameter.}
\end{figure}

In order for a numerical study, we set $M_X = 800$~GeV and $M_A =
173$~GeV, while $B_i$ is massless.
The invariant masses of visible particles $m_i$ are not vanishing, but
are set to be varying event-by-event between $0$ and
$153$~GeV.\footnote{
  In Sec.~\ref{sec:heavy_higgs},
  $m_i$ corresponds to the invariant mass of the $b$ quark and charged
  lepton system $m_{b\ell}$. The reader will notice that $0 \leq m_{b
    \ell} \leq \sqrt{m_t^2 - m_W^2} \simeq 153$~GeV, ignoring
  the $b$-quark mass.
}
These numbers have been chosen to match those used in the study of the
next section.
We also assume that the decay widths of the resonance and
intermediate particles are negligible.
In the left panel of Fig.~\ref{fig:at_ps}, we show the
$\Delta_\text{AT}$ distribution for pure phase space, ignoring the
spins and parities of all the particles involved in the decay process.
In order to see the impact of the singular feature on the
masses, we have used the true spatial momentum of the resonance, while
the mass $M_X$ is taken to be an input parameter. In other words, we
consider the case where the resonance was produced at rest.
For $M_X = M_X^\text{true}$, we can see a remarkably sharp peak at
$\Delta_\text{AT} = 0$, which confirms that the singular feature is
maximized at the true mass values.
Note that $\Delta_\text{AT}$ is negative since $\det (J J^\mathsf{T})
> 0$, while $\det \eta < 0$ for the correct choice of the mass
parameters.
Meanwhile, the distributions for $M_X \neq M_X^\text{true}$ do not
exhibit such a sharp peak: the event number densities around
$\Delta_\text{AT} = 0$ are not particularly noteworthy.

The observation suggests that the solutions of the singularity variable
$\Delta_\text{AT}$ have a direct and strong correlation with the
resonance mass.
We call the solutions $M_\text{AT}$,\footnote{In
  Ref.~\cite{Matchev:2019bon}, $M_X$ and $M_B$ are unknown, while
  $M_A$ is to be determined. Then, $M_\text{antler}$ as the solutions
  to the singularity variable has been introduced.
  We think that our consideration is more common in new physics
  searches at hadron colliders.
  Furthermore, as can be seen in Sec.~\ref{sec:heavy_higgs}, we
  perform a more realistic study beyond the phase space as well.}
\begin{equation}
  \Delta_\text{AT} \left( M_\text{AT};\, M_A, \, M_B,\, Q_L\right)
  = 0.
  \label{eq:AT_eq}
\end{equation}
Although there exist general analytic solutions to the polynomial
equations up to fourth order, it does not give us further insight, nor
particularly useful.
For practical use of the $\Delta_\text{AT}$ and $M_\text{AT}$
variables, we provide the coded implementations in
Haskell~\cite{Park:mantler} and \CC~\cite{Park:MAT}.
As mentioned earlier, since $\Delta_\text{AT}$ is essentially a
quadratic polynomial equation if $X$ is at rest, there are up to two
degenerate solutions to the equation.
We sort the solutions into the smaller and the larger, and label them
as $M_\text{AT}^\text{min}$ and $M_\text{AT}^\text{max}$, respectively.
The $M_\text{AT}$ distributions are shown in the right panel of
Fig.~\ref{fig:at_ps}.
Note that the two distributions overlap only at the true $M_X$ value.
It means that if a unique solution for Eq.~(\ref{eq:AT_eq}) exists for
given events, it corresponds precisely to the resonance mass.
One can see that the singularity of $\Delta_\text{AT}$ has transformed
into the peak of the $M_\text{AT}$ distribution at $M_X^\text{true}$, and
both $M_\text{AT}^\text{min}$ and $M_\text{AT}^\text{max}$ contribute
to the peak.
Moreover, the non-overlapping of the two distributions when
$M_\text{AT} \neq M_X^\text{true}$ leads us to conclude that
\begin{equation}
  M_\text{AT}^\text{min} \leq M_X^\text{true} \leq
  M_\text{AT}^\text{max} .\label{eq:at_min_max}
\end{equation}
This relation implies that we can directly deduce the mass scale of
$M_X$ from the edge of $M_\text{AT}^\text{min}$ and the threshold of
$M_\text{AT}^\text{max}$ distributions.
In practical applications, we expect that both variables can serve as
important cuts to enhance the signal-to-background ratio, in
particular at the stage of discovery.
We will see this aspect more clearly in Sec.~\ref{sec:heavy_higgs}
with a more concrete example in the presence of background.

Up to now, we have assumed that the longitudinal momentum of the
resonance is known {\em a priori}. However, it is practically
unattainable for hadron collider events.
In practice, the resonance is boosted in the longitudinal direction
by the net momentum sum of colliding partons, which follows parton
distribution functions.
The amount of the longitudinal boost is unknown on an event-by-event
basis.
It is an intrinsic nature of hadron colliders, one of the biggest
obstacles in particle object reconstructions.
In Refs.~\cite{Matchev:2019bon, DeRujula:2012ns}, the longitudinal
momentum of resonance has been neglected by assuming that the
colliding gluons have nearly the same amount of energy.
In our numerical simulation with parton distribution functions, we
found that the longitudinal momentum could be so large that it could
not be neglected, even in the case of gluon-gluon collision.
However, in the absence of a good estimator or approximator for the
unknown longitudinal momentum, we have no other choice but to take an
ad hoc solution. We ignore it by setting it to be zero, {\em {\`a} la}
transverse mass variables.
We denote the ad hoc solution as $M_\text{AT}^{(0)}$ to distinguish it
from the original definition,
\begin{equation}
  \Delta_\text{AT}^{(0)} \equiv \Delta_\text{AT} (
    M_\text{AT}^{(0)};\, M_A,\, M_B, \, Q_L = 0 ) .
\end{equation}
The corresponding $\Delta_\text{AT}^{(0)}$ is still essentially a
quadratic polynomial.

\begin{figure}[tb!]
  \begin{center}
    \includegraphics[width=0.48\textwidth]{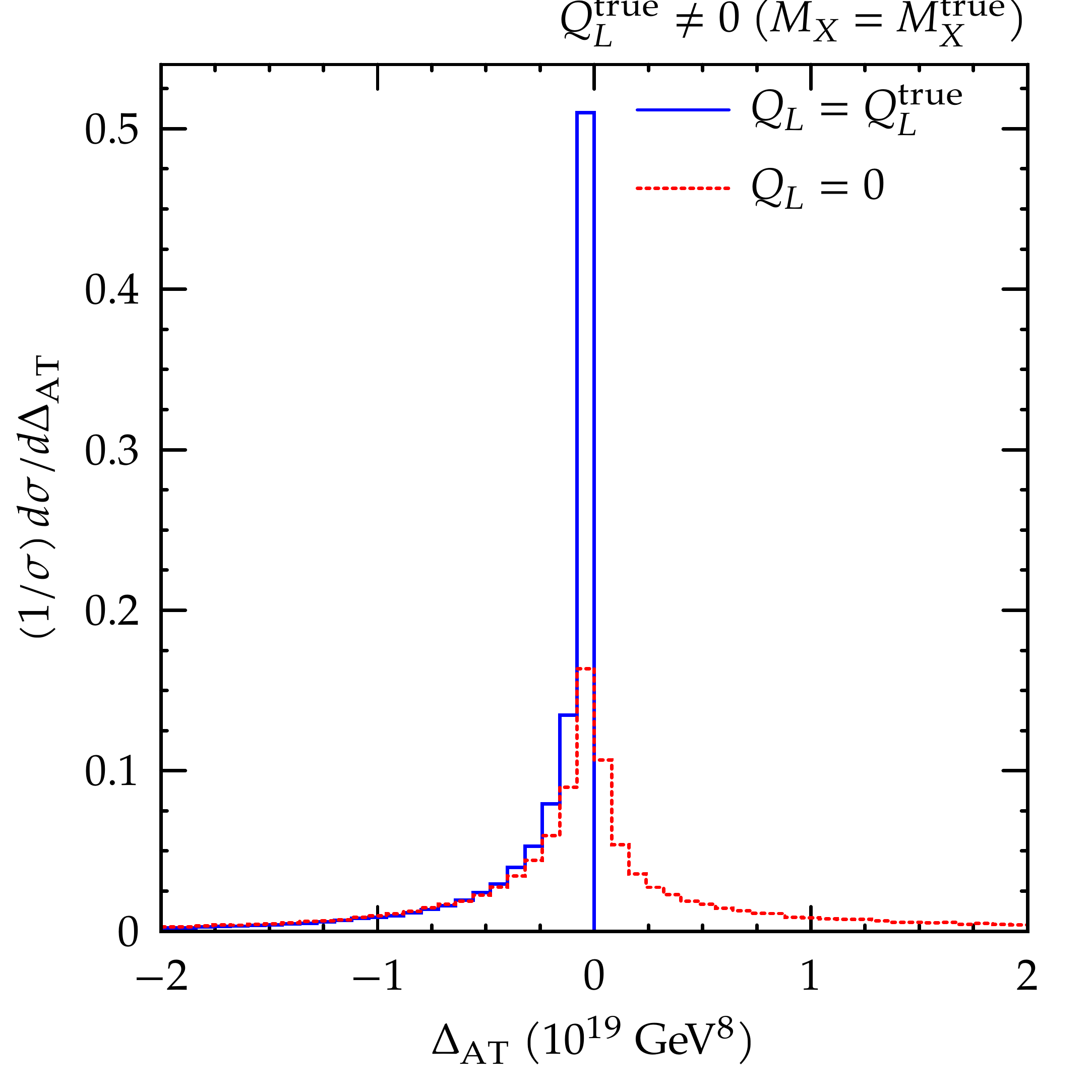}
    \includegraphics[width=0.48\textwidth]{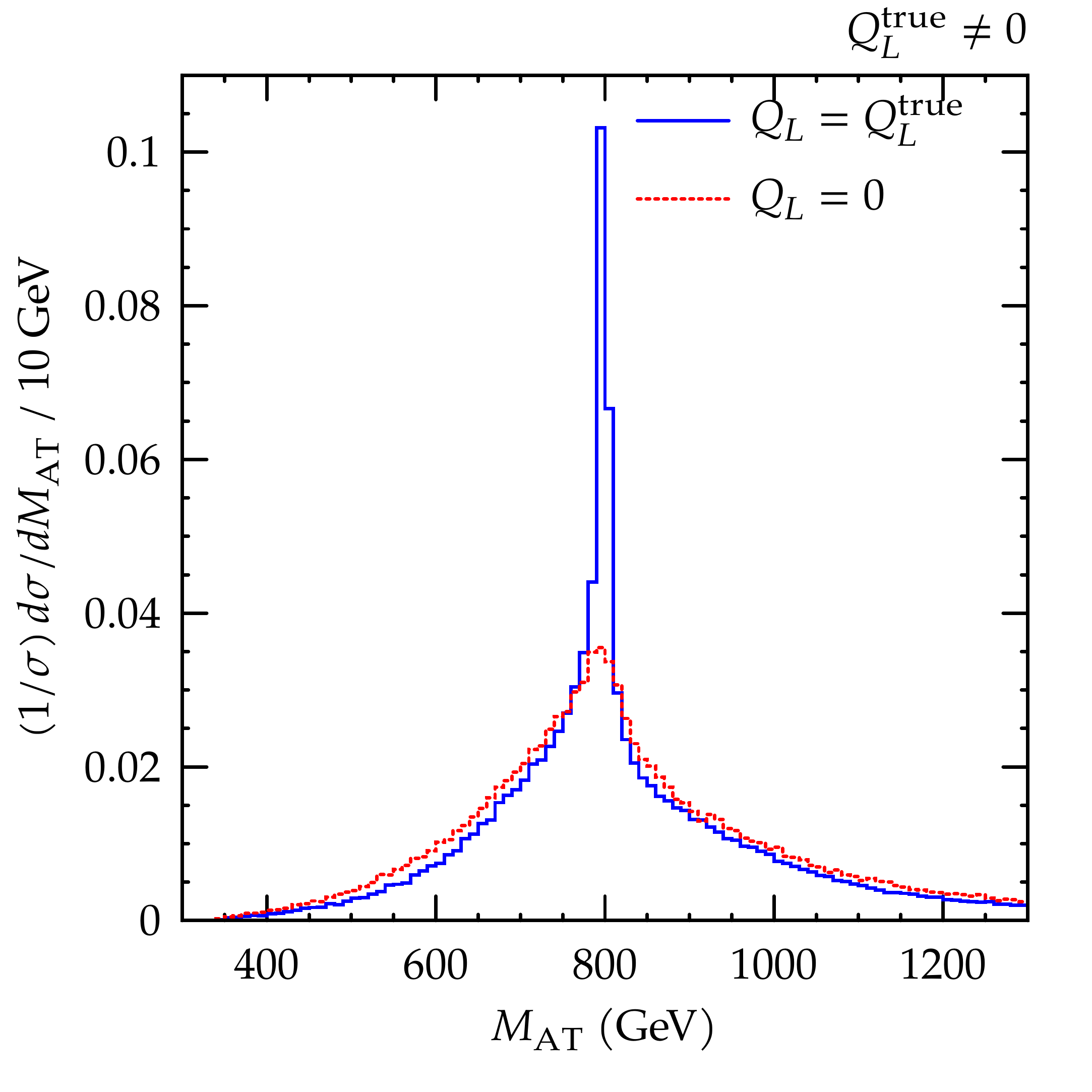}
    \includegraphics[width=0.48\textwidth]{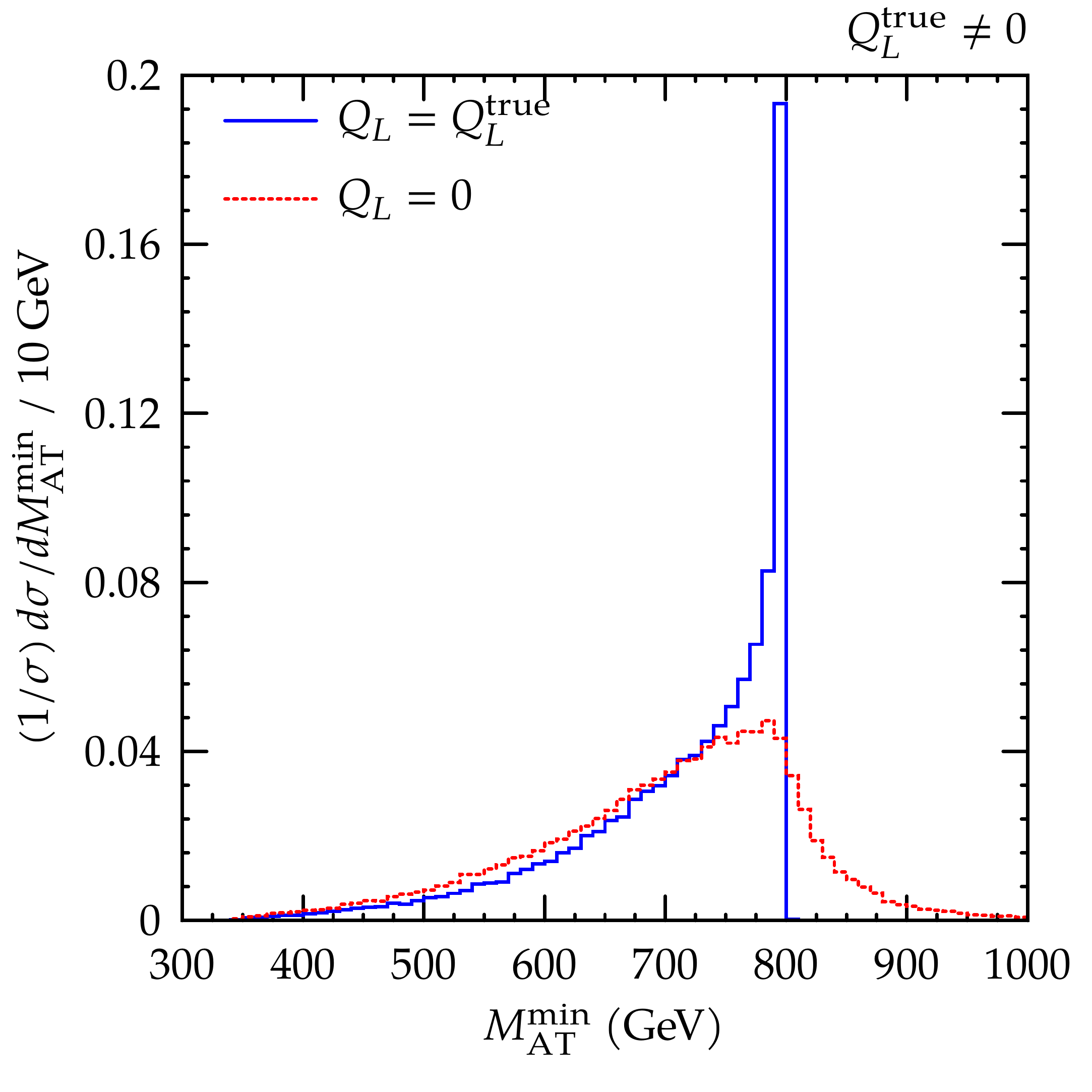}
    \includegraphics[width=0.48\textwidth]{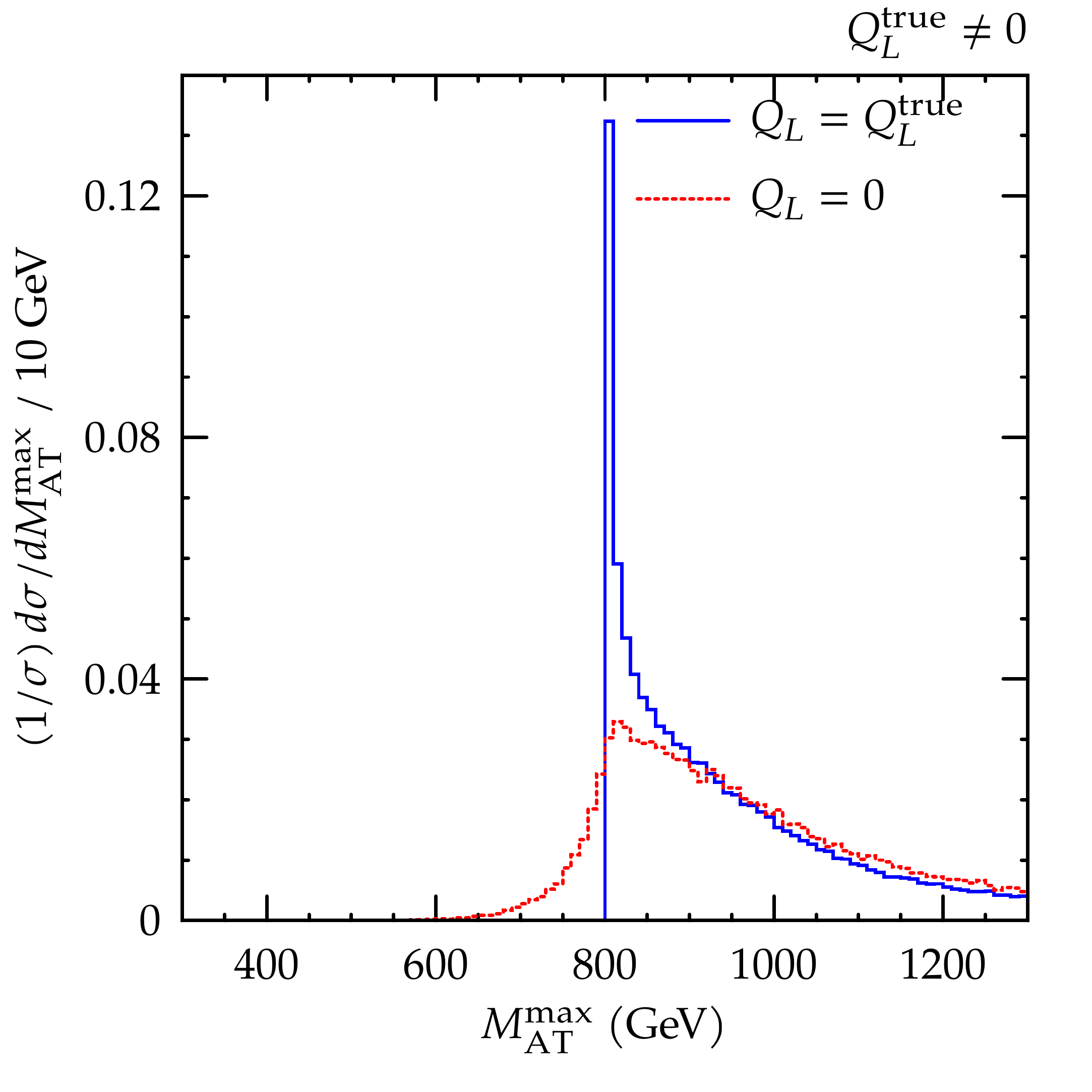}
  \end{center}
  \caption{\label{fig:at_ps_isr}
    The $\Delta_\text{AT}$ and $M_\text{AT}$ distributions in the
    presence of longitudinal boost of the resonance.
    In each panel, the blue distribution is obtained by assuming that
    the longitudinal momentum $Q_L$ is known, while the red one is by
    setting $Q_L = 0$ when computing the variables. The latter
    corresponds to $\Delta_\text{AT}^{(0)}$ or $M_\text{AT}^{(0)}$.}
\end{figure}
In Fig.~\ref{fig:at_ps_isr}, we have shown the
$\Delta_\text{AT}^{(0)}$ and $M_\text{AT}^{(0)}$ distributions.\footnote{
  To simulate the distribution of the longitudinal momentum, we used a
  simple inverse power formula,
  \begin{equation*}
    f(Q_L) \propto Q_L \left( 1 + Q_L / Q_\ast \right)^{-n}
  \end{equation*}
  with $Q_\ast = 100$~GeV and $n = 5$.
  The formula is inspired by the one used in the measurements of
  transverse momentum distributions of strange
  mesons~\cite{Aamodt:2011zza}.
  The formula was used for illustration purposes only. We will
  properly use the parton distribution function in the next section.
}
For comparison, the corresponding distributions obtained by using the
true longitudinal momentum $Q_L^\text{true}$ have also been added
to each panel.
By ignoring the longitudinal momentum, the relation
(\ref{eq:at_min_max}) is no longer valid.
The endpoint of the $M_\text{AT}^{(0), \text{min}}$
($M_\text{AT}^{(0), \text{max}}$) distribution in
Fig.~\ref{fig:at_ps_isr} lies slightly above (below) the resonance
mass value.
And, the $M_\text{AT}^{(0),\text{min}}$ and $M_\text{AT}^{(0),\text{max}}$ distributions
show a peak at the resonance mass, while they are smeared.
From these observations, we expect that the $M_\text{AT}^{(0)}$
variable can still give us a hint of the mass scale of the resonance
at the stage of bump hunting.
If a good estimator for the longitudinal momentum is supplemented, the
relation (\ref{eq:at_min_max}) can be approximately restored so that the
endpoint and peak shapes of the $M_\text{AT}$ distributions
become more pronounced.
We will employ one example of such an estimator in the next section.

\section{\label{sec:heavy_higgs}
  Searching for heavy Higgs bosons decaying into a top pair}

\noindent
As an application of the $M_\text{AT}$ variable, we consider heavy
neutral Higgs bosons in the 2HDM\@.
The current results on the SM Higgs measurements~\cite{Sirunyan:2018koj,
  Aad:2019mbh} favor the alignment limit, where one of the neutral
Higgs bosons has the SM-like couplings to the SM vector bosons.
In the alignment limit, the non-SM-like Higgs bosons $H$, $A$, and
$H^+$ interact among themselves more strongly, while the branching
ratios of $H \to WW$, $ZZ$, and $hh$ are all suppressed.
Moreover, if the heavy Higgs bosons are mass-degenerate as in
the decoupling limit of the supersymmetric model, the dominant decay
mode is $H \to t \bar t$ unless the ratio of Higgs vacuum expectation
values $\tan\beta$ is very large in the type II model.
For a review on the heavy Higgs boson decays in the alignment limit of
the 2HDM, see Ref.~\cite{Grzadkowski:2018ohf} and references therein.

Though the top-pair process is an important channel to search for the
heavy Higgs boson, it must overcome the SM $t \bar t$ background,
which has a huge cross-section and possesses exactly the same final
state as the Higgs signal.
Furthermore, the di-leptonic final state $2b + 2\ell + \slashed{E}_T$
contains two neutrinos that prevent from reconstructing the
center-of-mass frame of the Higgs boson.
The ATLAS collaboration at the LHC is still focusing on the
fully-hadronic~\cite{Aad:2020kop} and semi-leptonic decay
modes~\cite{Aaboud:2018mjh}, but the CMS collaboration
has recently performed the search using the di-leptonic final
state~\cite{Sirunyan:2019wph}.
% In the CMS analysis, the invisible neutrino momenta have been obtained
% by directly solving the kinematic constraints, following the method in
% Ref.~\cite{Khachatryan:2015oqa}.
In the CMS analysis, the invisible neutrino momenta have been obtained
by following the method in Ref.~\cite{Khachatryan:2015oqa}.
They directly solve the series of equations for the kinematic
constraints and assign a weight to each solution to
characterize how likely it is to occur in $t \bar t$ production based
on the expected true $b$-lepton invariant mass spectrum.
Then, kinematic quantities, such as the invariant mass of the $t
\bar t$ system, are calculated as a weighted average.
We note that though the singularity variable is based on the same
kinematic constraints, we do not attempt to obtain the invisible
neutrino momenta.
Instead of solving all the kinematic constraints, we introduce the
singularity condition as given in~(\ref{eq:AT_eq}) and solve only the
one polynomial equation.
Therefore, it is computationally cheaper than the kinematic
reconstruction method used in the CMS analysis.

For the numerical study, we have chosen a benchmark point: the
CP-conserving type II model with $M_H = M_A = M_{H^+} = 800$~GeV,
$\tan\beta = 3$, and the Higgs mixing angle $\cos (\beta - \alpha) =
0.01$.
The benchmark point is safe from the existing LHC bounds set by the SM
Higgs decay width and coupling measurements and direct searches for
the heavy Higgs bosons~\cite{Kling:2020hmi}.
In our estimation, the branching ratio of the $H \to t \bar t$ process
is about $93$\%, while the subleading process is $H \to b \bar b$ with
the branching ratio of $\lesssim 3$\%.
The total decay width of the heavy Higgs boson is $2.7$~GeV.
We have generated parton-level event samples for both signal and
background using \texttt{Pythia 8}~\cite{Sjostrand:2014zea},
interfacing with \texttt{LHAPDF 6}~\cite{Buckley:2014ana} for parton
distribution functions.
We use the \texttt{NNPDF} parton distributions~\cite{Ball:2014uwa},
and set the proton-proton collision energy to be $13$~TeV.
In our estimation, we find that the gluon-fusion process dominates the
Higgs production, but the bottom-fusion process is non-negligible as
well.
The latter contributes to about $15\%$ of the total cross section at
leading order.
It is because of the enhanced Higgs coupling to the bottom quarks by
$\tan\beta$. In the case of the type I model, the bottom-fusion
process are suppressed, so the gluon-fusion contribution will be
predominant.

\begin{figure}[tb!]
  \begin{center}
    \includegraphics[width=0.48\textwidth]{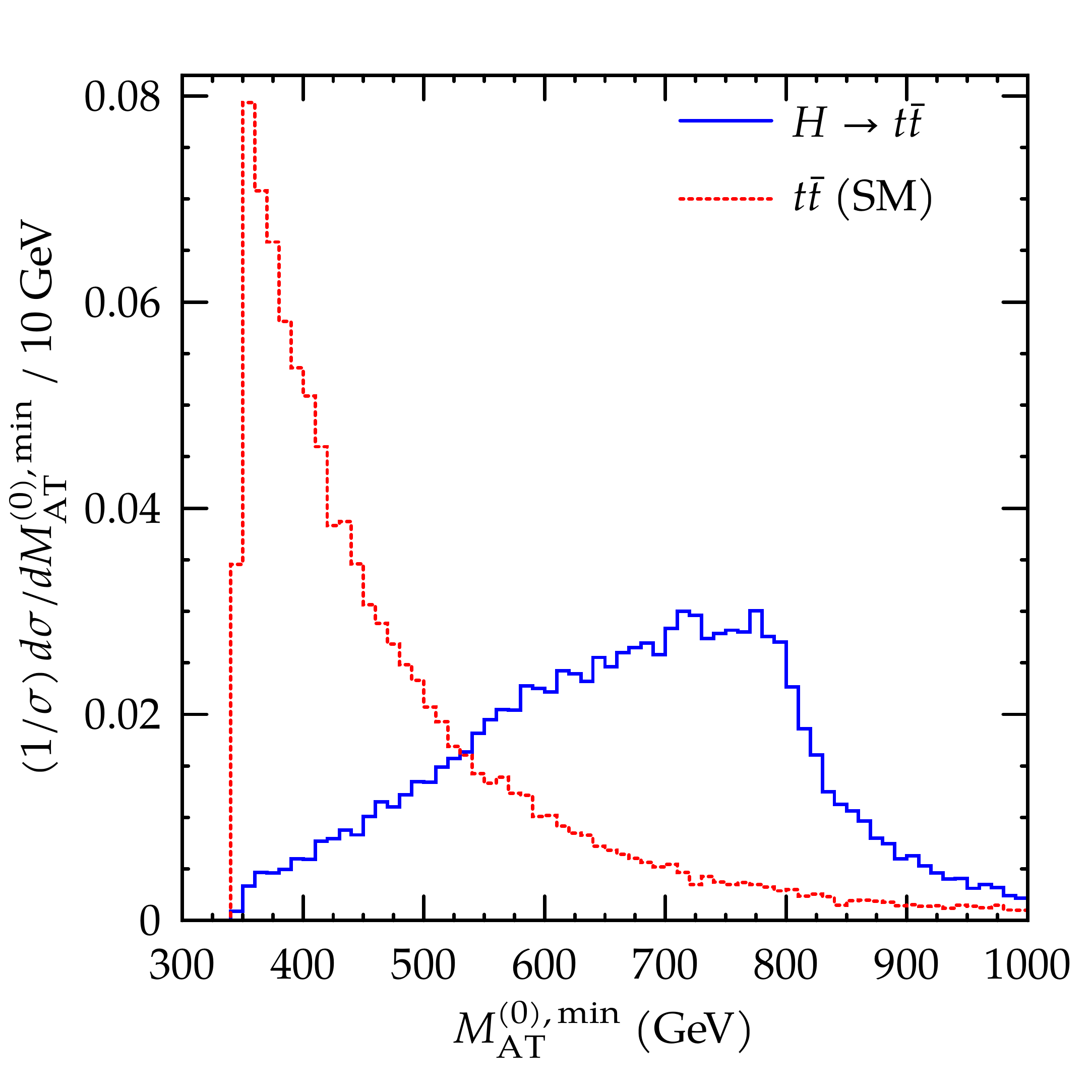}
    \includegraphics[width=0.48\textwidth]{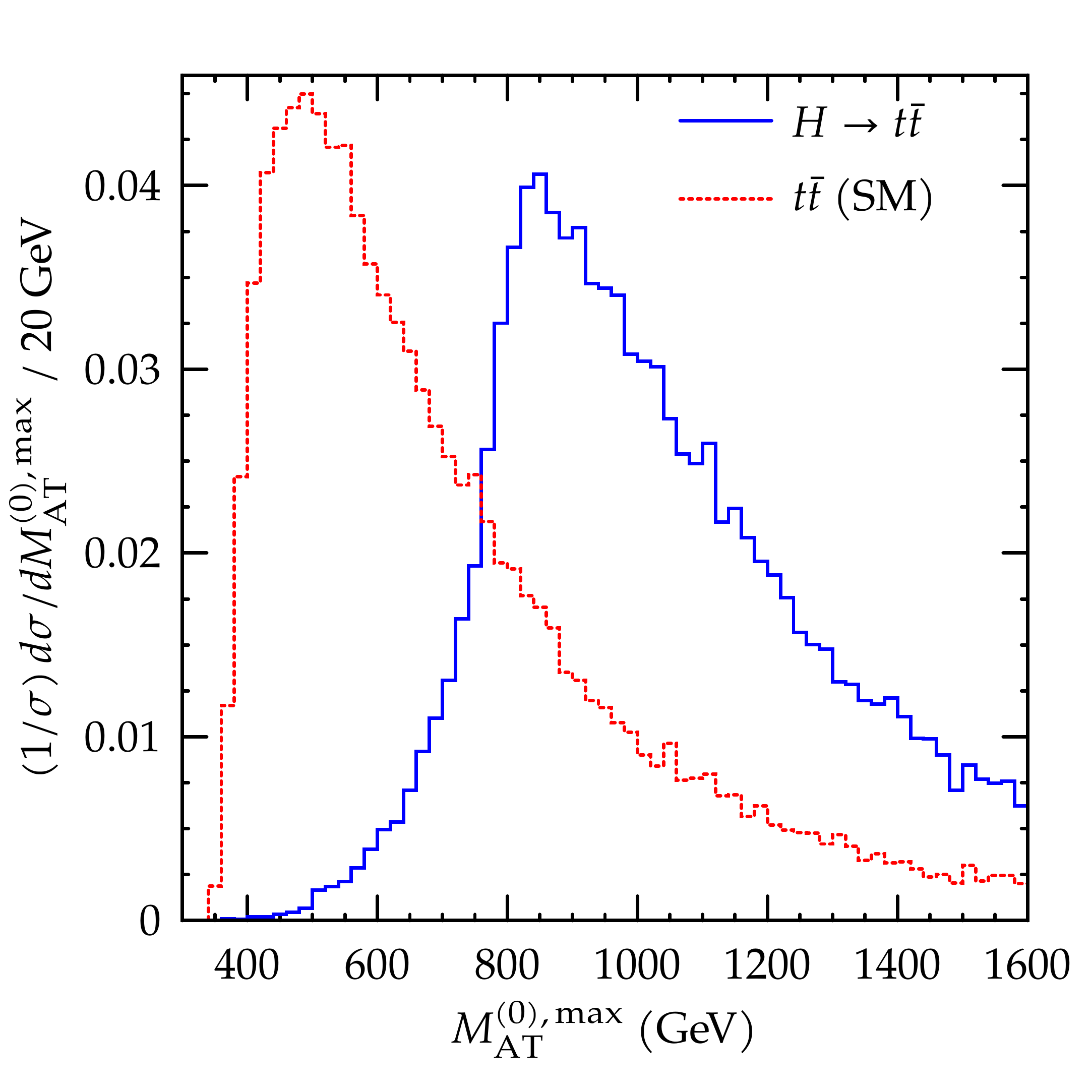}
  \end{center}
  \caption{\label{fig:mAT_parton}
    The $M_\text{AT}^{(0)}$ distributions for the $H \to t \bar t$
    signal and the SM $t \bar t$ background at parton level. We have
    used the correct pairing of the visible particles.}
\end{figure}

The parton-level distributions for the $M_\text{AT}^{(0)}$ variables
are displayed in Fig.~\ref{fig:mAT_parton}.
The signal and background distributions are well separated, and the
signal distributions have the endpoints near $M_H$.
For the $t \bar t$ background, $M_\text{AT}^{(0)}$ is correlated with
the invariant mass of the top pair $\sqrt{\hat s} = m_{t \bar t}$, so
the event number density is the largest near the threshold.
These observations show that the $M_\text{AT}^{(0)}$ variables can be
useful for the discovery of heavy Higgs bosons.
In practice, there is a combinatorial ambiguity of pairing the visible
particles into two sets. In Fig.~\ref{fig:mAT_parton}, we have used
the correct pairing. We will address this issue shortly.

As mentioned in the previous section, a good estimator for the
longitudinal momentum of the heavy Higgs boson can improve the
singularity variables.
It will be particularly in need when attempting the mass measurement
after discovery.
One possible option worth to examine is the $M_{T2}$-assisted on-shell
(MAOS) method~\cite{Cho:2008tj, Choi:2009hn, Choi:2010dw,
  Park:2011uz}.
The MAOS method provides an approximation for the longitudinal momenta
of invisible particles by solving the on-shell relations, given the
solution of the $M_{T2}$ variable for the transverse
momenta.\footnote{
  The longitudinal momenta can also be approximated by using $M_2$
  instead of $M_{T2}$~\cite{Cho:2014naa}.
  We expect that the accuracy would be of similar or better quality
  than the MAOS method.
}
The $M_{T2}$ variable can be used in the presence of two invisible
particles in the final state~\cite{Lester:1999tx, Barr:2003rg}.
The endpoint of the $M_{T2}$ distribution is $m_t$ for both signal and
background because the top quarks were produced on mass-shell.
From the two quadratic on-shell relations, the MAOS method yields
up to four possible solutions for the unknown longitudinal momenta.
By setting the longitudinal momentum of the resonance or the
$t \bar t$ system to be
\begin{equation}
  Q_L^\text{maos}
  = p_{1L} + p_{2L} + k_{1L}^\text{maos} + k_{2L}^\text{maos} ,
\end{equation}
the singularity condition becomes the quartic polynomial of
$Q_0^\text{maos} =$ $(M_\text{AT}^2 +$ $\norm{\vb*{Q}_T}^2 +$
$(Q_L^\text{maos})^2)^{1/2}$,
\begin{equation}
  \Delta_\text{AT} (Q_0^\text{maos};\, M_A, \, M_B) = 0 .
    \label{eq:delta_AT_maos}
\end{equation}
From the solution of the polynomial equation in the above, we have
\begin{equation}
  M_\text{AT}^\text{maos} = \sqrt{(Q_0^\text{maos})^2 -
    \norm{\vb*{Q}_T}^2 - (Q_L^\text{maos})^2} .
\end{equation}
Note that the total number of $M_\text{AT}^\text{maos}$ for given
event is now up to $16$.
% In the numerical study, we find that about half of them are not real.
As we find no plausible criterion to choose a particular solution, we
shall use all the real solutions for $M_\text{AT}$, discarding the complex
ones.
As the number of solutions has been increased, we define
$M_\text{AT}^\text{maos,min}$ and $M_\text{AT}^\text{maos,max}$ to
be the smallest and largest solution of Eq.~(\ref{eq:delta_AT_maos}),
respectively.

\begin{figure}[tb!]
  \begin{center}
    \includegraphics[width=0.48\textwidth]{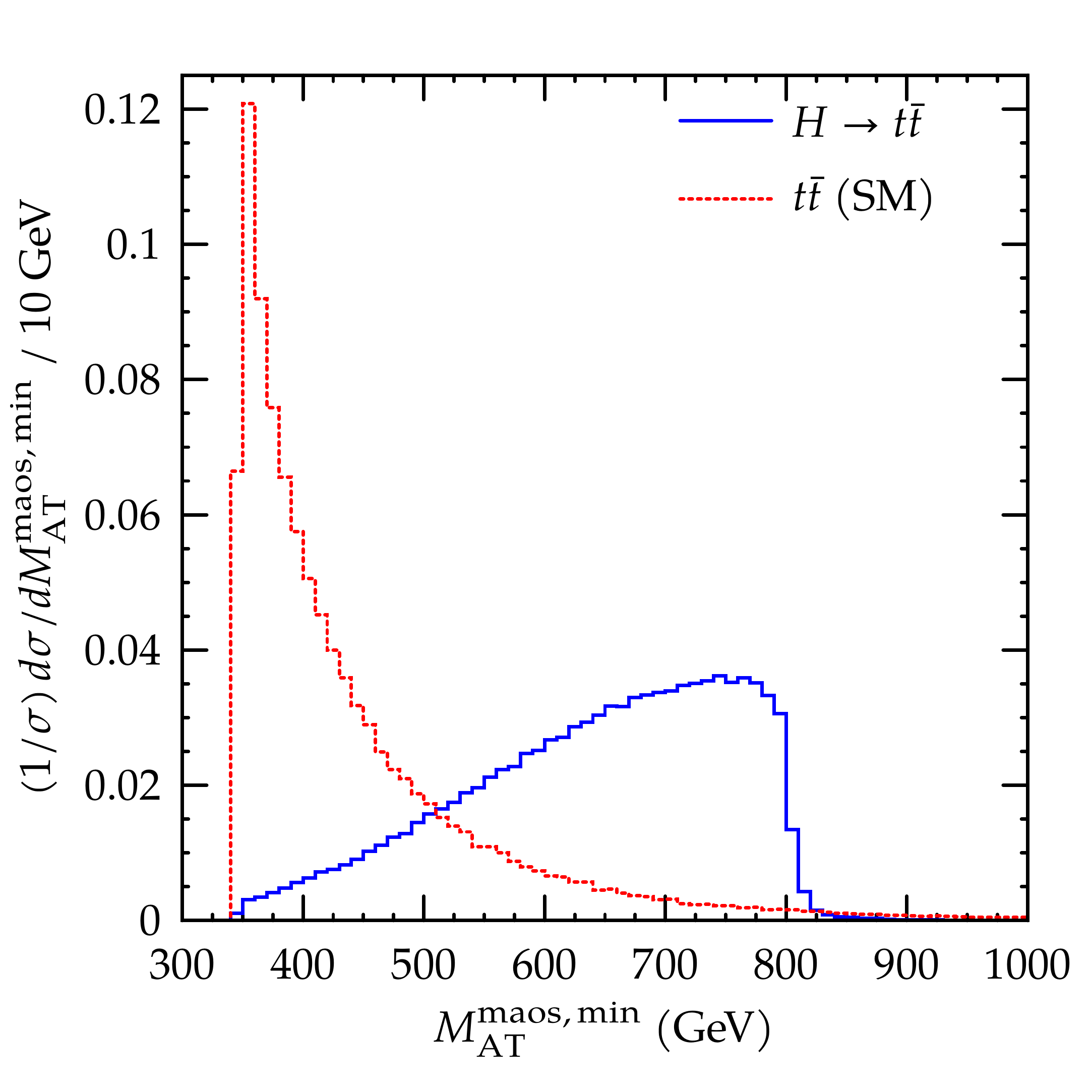}
    \includegraphics[width=0.48\textwidth]{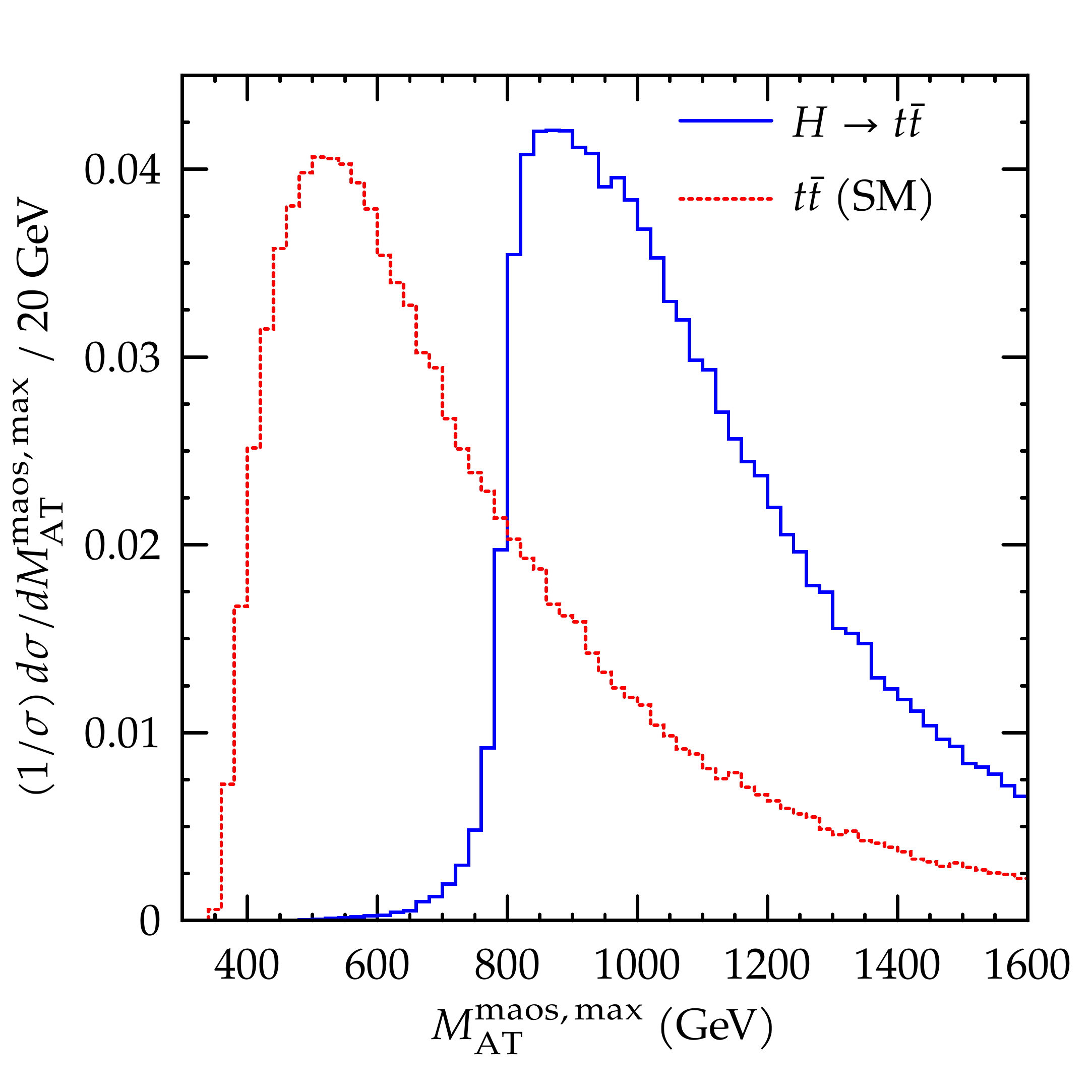}
  \end{center}
  \caption{\label{fig:mAT_maos_parton}
    The $M_\text{AT}^\text{maos}$ distributions for the $H \to t \bar t$
    signal and the SM $t \bar t$ background at parton level. The MAOS
    method has been used to obtain the approximation for the
    longitudinal momentum of the $t \bar t$ system.
    We have used the correct pairing of the visible particles.
  }
\end{figure}

We show the $M_\text{AT}$ distributions using the MAOS method in
Fig.~\ref{fig:mAT_maos_parton}.
One can see that the endpoints of the $M_\text{AT}$ distributions are
more pronounced than $M_\text{AT}^{(0)}$, and lie at the heavy Higgs
boson mass.
It is known that the accuracy of the longitudinal momentum
approximation can be improved by imposing an $M_{T2}$
cut~\cite{Cho:2008tj}, so it can make the endpoint shape more
distinct.
Therefore, we expect that the $M_\text{AT}$ with the MAOS method can
serve as a useful variable to measure the resonance mass accurately
with the $M_{T2}$ cut.

We here briefly leave a comment on the combinatorial ambiguity on the
pairing of visible particles. There are two possible ways to pair one
$b$ quark and one charged lepton in each event.
One can resolve the ambiguity by using various kinematic variables.
In particular, the algorithm proposed in Ref.~\cite{Choi:2011ys} and
the improved version~\cite{Debnath:2017ktz} provide a good efficiency
$\gtrsim 80\%$ for $t \bar t$ events.
In practice, the combinatorial ambiguity does not interfere with
the endpoints of $M_\text{AT}^\text{min}$ and $M_\text{AT}^\text{max}$
distributions.
One computes $M_\text{AT}$ for all possible pairing and then take the
minimum or maximum. The positions of the endpoints are intact by
taking the extrema.
Nonetheless, we have checked that the algorithms work well for both
signal and background events, and the combinatorial ambiguity does not
affect the overall shape of the $M_\text{AT}$ distributions when using
the algorithms.

Finally, we examine the $M_\text{AT}$ variable including jet
reconstruction, object isolation, event selection cuts, and various
detector effects.
The parton-level event samples have been processed by \texttt{Pythia}
for parton shower and hadronization.
Then, we employ \texttt{FastJet 3} for reconstructing
jets~\cite{Cacciari:2011ma}. In our simulation, the anti-$k_T$
clustering algorithm~\cite{Cacciari:2008gp} with a distance parameter
$R = 0.4$ is chosen for the jet reconstruction.
The object reconstructions and detector effects, such as flavor
tagging, fake rates, and momentum smearing, have been performed by
the fast detector simulation program \texttt{DELPHES
  3}~\cite{deFavereau:2013fsa}.
We set the cone sizes for isolating the electrons and muons to be
$0.4$. The jet energy scale correction is fixed to unity.
Except for the settings described in the above, we use the default CMS
card provided by \texttt{DELPHES}.

\begin{figure}[tb!]
  \begin{center}
    \includegraphics[width=0.48\textwidth]{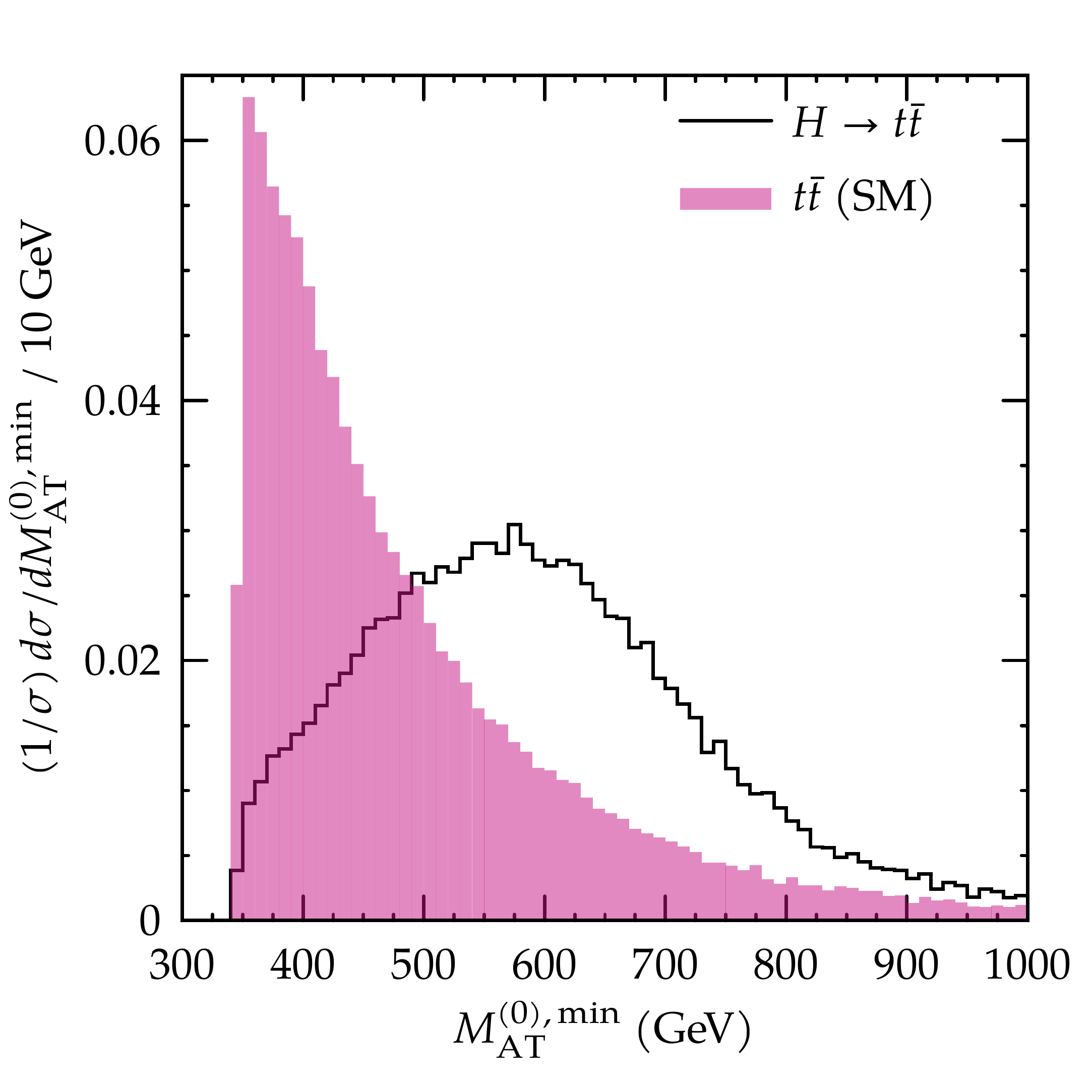}
    \includegraphics[width=0.48\textwidth]{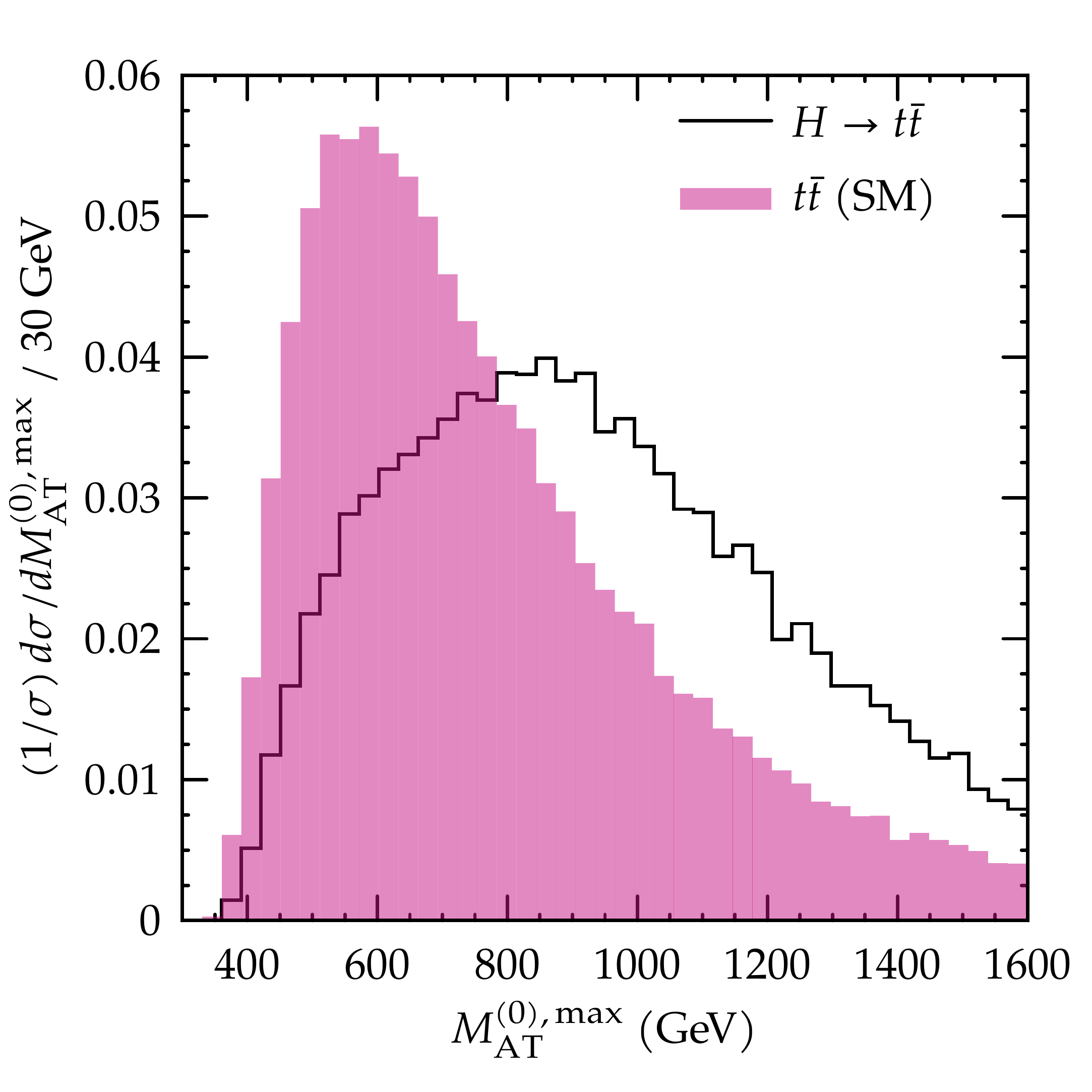}
    \includegraphics[width=0.48\textwidth]{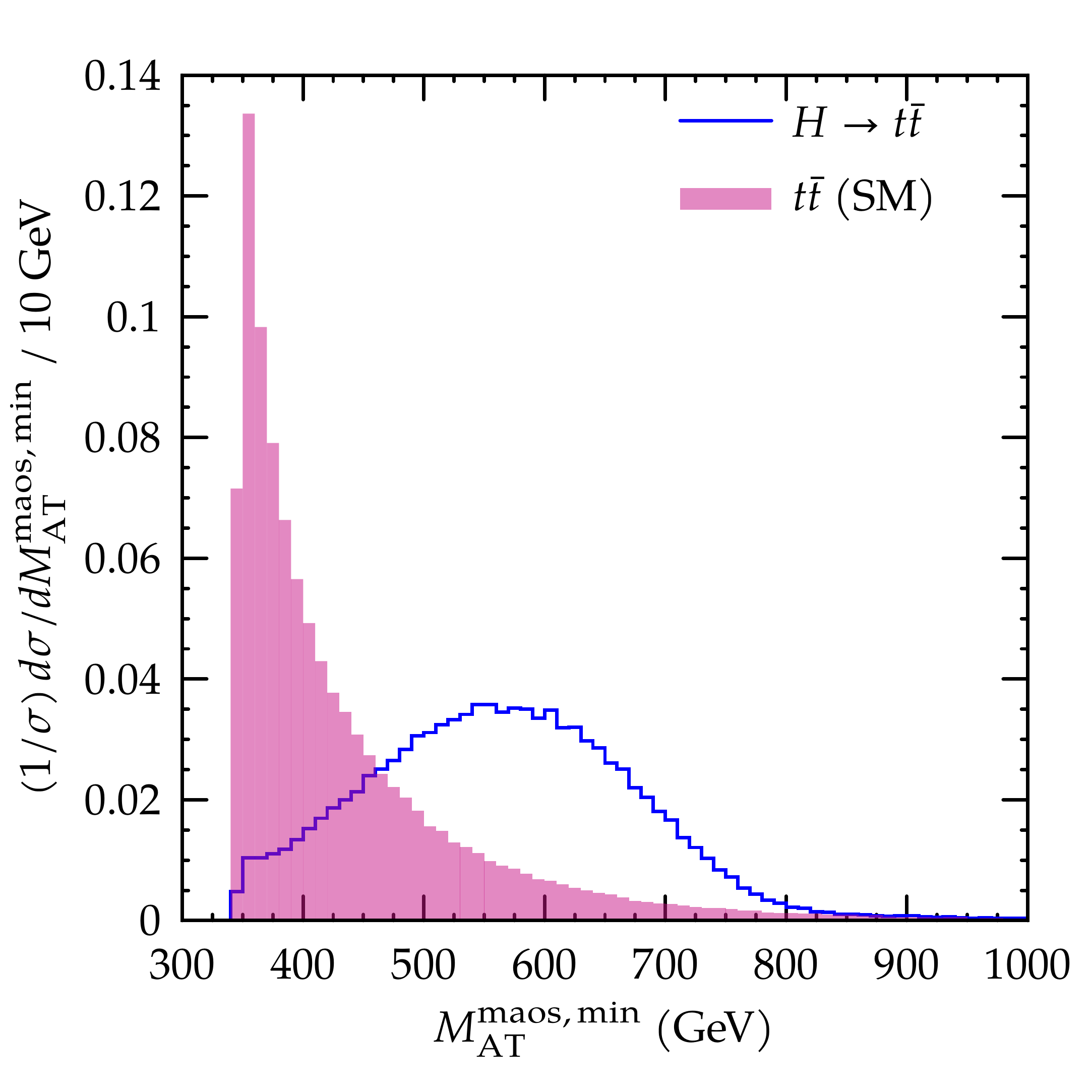}
    \includegraphics[width=0.48\textwidth]{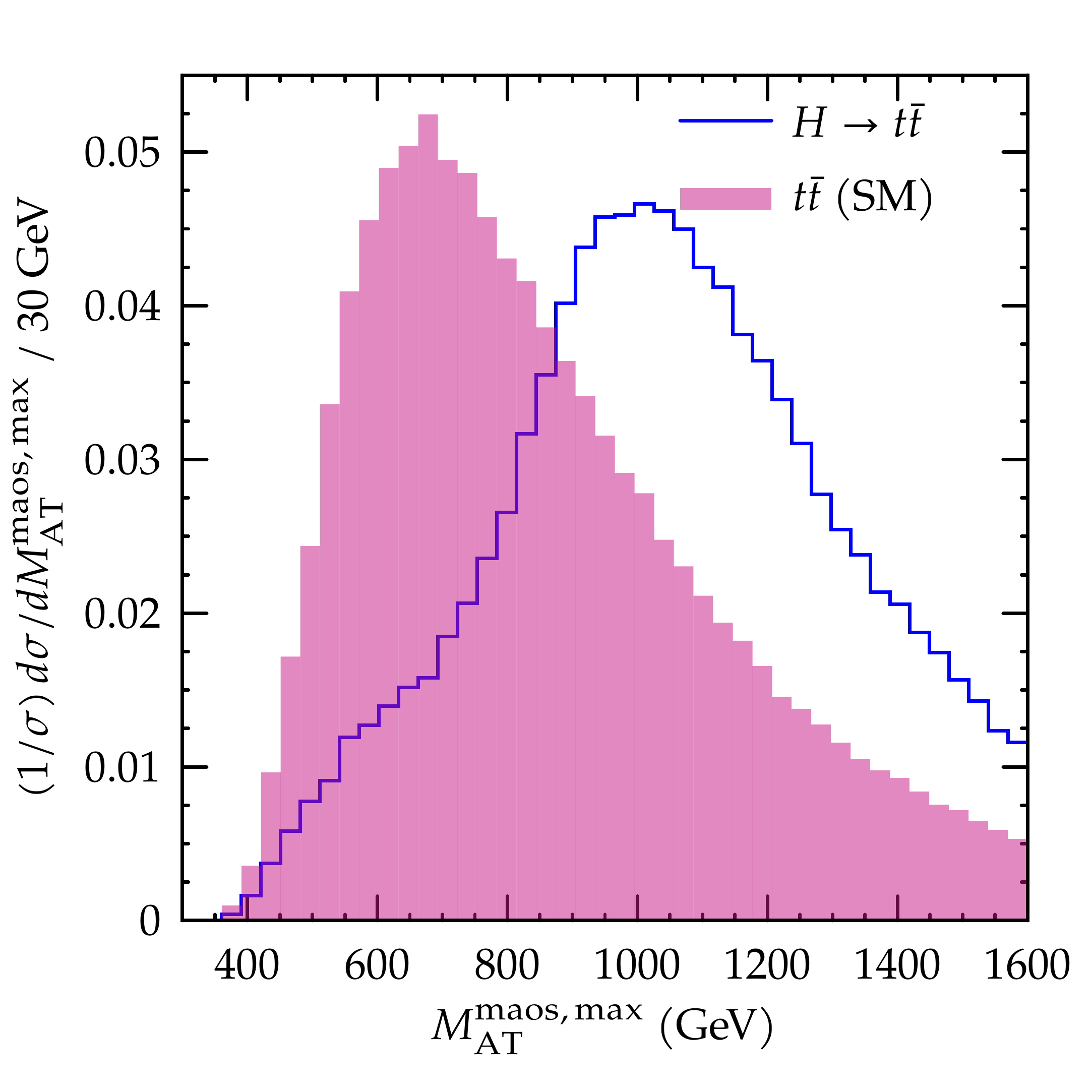}
  \end{center}
  \caption{\label{fig:mAT_lhc}
    The $M_\text{AT}^{(0)}$ (upper) and $M_\text{AT}^\text{maos}$
    (lower) distributions for detector-level events.
    We have applied the basic selection cuts.
  }
\end{figure}

We have applied the basic selection cuts similar to those used in the
CMS analysis to the detector-level events~\cite{Sirunyan:2019wph}:
\begin{itemize}
\item Isolated electrons and muons satisfy $p_T > 20$~GeV and
  $|\eta| < 2.4$. Only events with two oppositely charged leptons are
  selected. The leading lepton must have $p_T > 25$~GeV.
\item The invariant mass of isolated leptons is required to be larger
  than $20$~GeV to suppress low-mass resonance events. And, in order
  to veto events with the $Z$ boson, we reject events with $76 <
  m_{\ell \ell} < 106$~GeV.
\item We require that at least two jets with $p_T > 30$~GeV and
  $|\eta| < 2.4$, and at least one of the jets are $b$ tagged.
\item The missing energy is required to be larger than $40$~GeV.
\end{itemize}

In Fig.~\ref{fig:mAT_lhc}, we show the $M_\text{AT}$ distributions for
the detector-level event data.
Here we have used the algorithm in~\cite{Choi:2011ys} to resolve the
combinatorial ambiguity.
In the $M_\text{AT}^\text{min}$ distributions, the edge structure is
clearly visible around the $m_H$ value, and the background
distributions are populated near the threshold.
Meanwhile, the peak has been shifted towards below $m_H$.
For $M_\text{AT}^\text{max}$, the peak is located at slightly above
the $m_H$ value.
One of the possible reasons for the peak shift is that some of the
decay products of the $b$ quark could be missed since $b$ jets tend to
be wider and have higher multiplicities than light jets.
Consequently, the $b$ jet
may not catch the momentum of the initiating $b$ quark well enough.
It can be improved by investigating flavor-dependent jet energy scale
corrections and comparing various jet reconstruction algorithms and
the inherent parameters associated with the algorithms.
On the other hand, we find that the momentum smearing effect and
basic selection cuts do not distort much the overall shape of the
distributions.
It is worth scrutinizing the variables in more detail at jet level,
but it is beyond the scope of our study.
We note that the $M_\text{AT}$ variables are not affected by initial
state radiations since the variables are derived from the Lorentz-invariant
kinematic constraints, given in~(\ref{eq:kinematic_constraints}),
in which the radiations do not participate.

The simulation results show that one can deduce the heavy Higgs mass
by investigating the $M_\text{AT}^\text{min}$ and
$M_\text{AT}^\text{max}$ distributions.
The $M_\text{AT}^\text{min}$ distribution has the edge structure at
the heavy Higgs mass, even when taking into account the jet
reconstruction and detector effects.
The MAOS approximation method for the longitudinal momentum of the
heavy Higgs boson makes the edge structure more clearer.
One can see that the MAOS method has improved the shapes of the
$M_\text{AT}$ distributions by comparing the upper and the lower
panels in Fig.~\ref{fig:mAT_lhc}.
The threshold of the signal $M_\text{AT}^\text{min}$ distribution can
be buried by the background distribution, but the peak of the
distribution may give us the information of the upper bound on the
heavy Higgs mass.
Furthermore, the signal distributions remain to be well separated from the
background.
This observation suggests that the $M_\text{AT}$ variable can also be
useful for setting the signal or control regions when searching for
heavy resonances.

\section{Conclusions}

\noindent
The algebraic singularity method was proposed from the observation
that the projected visible phase space can have singularities in the
presence of missing energy.
The singularity variables have been devised to capture such singular
features, and they implicitly provide the mass spectrum information of
intermediate resonances and invisible particles in the final state.
Recently, it has been outlined the prescriptions for deriving the
singularity variables for various event topology with missing
energy~\cite{Matchev:2019bon}.
It makes the singularity method more accessible for practical
applications.

In this article, we focused on the antler decay topology, where a
heavy resonance decays into the final state of two visible and two
invisible particles through intermediate states.
We have identified the singularity condition using the prescription in
Ref.~\cite{Matchev:2019bon} and derived a one-dimensional variable
that has a strong correlation with the resonance mass.
The $M_\text{AT}$ variable is the solution to the polynomial equation
of the singularity condition.
We have confirmed that the phase-space distributions of the minimum
and maximum of the $M_\text{AT}$ have endpoints at the correct
resonance mass value, thus enabling us to measure the mass as well as
to discover the resonant signal.

As a practical application, we have studied the signal of the heavy
Higgs bosons decaying into a top pair, one of the typical signals in
the 2HDM\@.
Although the ignorance of the longitudinal momentum of the heavy Higgs
boson smears the endpoint structure of the $M_\text{AT}$
distributions, we find that the signal distributions are well
separated from the background, and they can give us a hint for the
mass scale of the heavy Higgs boson.
Moreover, by employing an approximation scheme for the longitudinal
momentum, the endpoint structure can become sharper.
We have used the MAOS method to exemplify our proposal and found that
it successfully restores the shape of the $M_\text{AT}$ distributions.
The feature remains intact even in the presence of the detector
effects and selection cuts, on the whole.
From these observations,
we expect that the $M_\text{AT}$ can serve as
the main variable of cut-based analyses, or an important input
feature of multivariate studies, for heavy resonance searches at
hadron colliders.

\section*{Acknowledgments}

We hope people all over the world stay strong and healthy with their
loved ones amid pandemic.
The author would like to thank Yeong~Gyun~Kim and Seodong~Shin for
useful comments on the manuscript.
This work was supported by IBS under the project code, IBS-R018-D1.

\bibliography{mat}

\end{document}